# Learning Expressive Linkage Rules using Genetic Programming

Robert Isele
Web-based Systems Group
Freie Universität Berlin
Garystr. 21, 14195 Berlin, Germany
mail@robertisele.com

Christian Bizer
Web-based Systems Group
Freie Universität Berlin
Garystr. 21, 14195 Berlin, Germany
chris@bizer.de

## ABSTRACT

A central problem in data integration and data cleansing is to find entities in different data sources that describe the same real-world object. Many existing methods for identifying such entities rely on explicit linkage rules which specify the conditions that entities must fulfill in order to be considered to describe the same real-world object. In this paper, we present the GenLink algorithm for learning expressive linkage rules from a set of existing reference links using genetic programming. The algorithm is capable of generating linkage rules which select discriminative properties for comparison, apply chains of data transformations to normalize property values, choose appropriate distance measures and thresholds and combine the results of multiple comparisons using non-linear aggregation functions. Our experiments show that the GenLink algorithm outperforms the state-of-the-art genetic programming approach to learning linkage rules recently presented by Carvalho et. al. and is capable of learning linkage rules which achieve a similar accuracy as human written rules for the same problem.

## 1. INTRODUCTION

As companies move to integrating data from even larger sets of internal and external data sources and as more and more structured data is becoming available on the public Web, the problem of finding entities in different data sources that describe the same real-world object is moving into the focus within even more application scenarios.

This problem has been studied extensively in the database community and is known as entity matching, record linkage, coreference resolution and deduplication [12, 16, 31].

Many existing methods identify matching entities using rule-based approaches [22]. Within these methods, linkage rules [31] specify the conditions that two entities must fulfill in order to be be considered to describe the same real-world object. Linkage rules typically compare different properties of the entities using a set of distance measures. The resulting similarity scores may be combined using different aggregation functions. If the data sources use different property value representation formats, property values may be normalized by applying transformations prior to the comparison. Writing good linkage rules by hand is a non-trivial problem as the rule author needs to have detailed knowledge about the source data set and the target data set in order to choose appropriate properties, data transformations, distance measures together with good thresholds as well as aggregation functions.

In this paper, we present *GenLink*, a supervised learning algorithm which employs genetic programming in order to learn linkage rules from a set of existing reference links. GenLink is capable of matching entities between heterogeneous data sets which adhere to different schemata. By employing an expressive linkage rule representation the algorithm learns rules which:

- Select discriminative properties for comparison.
- Apply chains of data transformations to normalize property values prior to comparison.
- Apply multiple distance measures combined with appropriate distance thresholds.
- Aggregate the result of multiple comparisons using linear as well as non-linear aggregation functions.

Following genetic programming, the GenLink algorithm starts with an initial population of candidate solutions which is iteratively evolved by applying a set of genetic operators. The basic idea of GenLink is to evolve the population by using a set of specialized crossover operators. Each of these operators only operates on one aspect of the linkage rule e.g. one crossover operator builds chains of transformations while another operator recombines different comparisons.

**Contributions.** In this paper, we make the following contributions: (1) We introduce a linkage rule representation which combines different distance measures non-linearly and may include chains of data transformations to normalize values prior to comparison. Our linkage rule representation is more expressive than previous work and subsumes threshold-based boolean classifiers and linear classifiers. Linkage Rules are represented as an operator tree and can be understood and further improved by humans. (2) We propose the GenLink genetic programming algorithm which uses specialized crossover operators to evolve linkage rules covering the full expressivity of the introduced representation. (3) We show that GenLink outperforms the state-of-the-art genetic programming approach for entity matching recently presented by Carvalho et. al. [10] and is capable of learning linkage rules which achieve a similar accuracy as human written rules for the same problem. (4) We have implemented GenLink as part of the Silk Link Discovery Framework [20]. Silk discovers matching entities within data sets that are represented as RDF. The main application area of the framework is to find matching entities within data sets that are accessible on the Web according to the Linked Data principles [4]. The Silk Link Discovery Framework[1] is available for download under the terms of the Apache License and all experiments that are presented in this paper can thus be repeated by the interested reader.

This paper builds on our previous work published as [19] and extends it with a formalization of the linkage rule representation, a more detailed description of the operators and a more extensive evaluation.

**Paper Organization.** The rest of the paper is organized as follows: Section 2 formalizes the entity matching problem. Based on that, Section 3 introduces our linkage rule representation. Section 4 discusses related work. Section 5 describes the GenLink algorithm in detail. Section 6 presents the results of the experimental evaluation.

## 2. PROBLEM DEFINITION

We consider the problem of matching entities between two data sources $A$ and $B$. Each entity $e \in A \cup B$ can be described with a set of properties $e.p_1, e.p_2, \ldots e.p_n$. For instance, an entity denoting a person may be described by the properties *name*, *birthday* and *address*. The objective is to determine which entities in $A$ and $B$ identify the same real world object.

The general problem of entity matching can be formalized as follows [15]:

DEFINITION 1 (ENTITY MATCHING). *Given two data sources $A$ and $B$, find the subset of all pairs of entities for which a relation $\sim_R$ holds:*

$$M = \{(a,b); a \sim_R b, a \in A, b \in B\}$$

*Similarly, we define the set of all pairs for which $\sim_R$ does not hold:*

$$U = (A \times B) \setminus M$$

The purpose of relation $\sim_R$ is to relate all entities which represent the same real world object.

In some cases a subset of $M$ and $U$ is already known prior to matching. Such *reference links* can, for instance, originate from previous data integration efforts. Alternatively, they can be created by domain experts who simply need to confirm or reject the equivalence of entity pairs from the data sets. Creating reference links is much easier than to write linkage rules as it requires no previous knowledge about similarity computation techniques or the specific linkage rule format used by the system. In [21], we present an active learning method to minimize the number of entity pairs which need to be confirmed or rejected.

DEFINITION 2 (REFERENCE LINKS). *A set of positive reference links $R_+ \subseteq M$ contains pairs of entities for which relation $\sim_R$ is known to hold (i.e. which identify the same real world object). Analogously, a set of negative reference links $R_- \subseteq U$ contains pairs of entities for which relation*

[1] http://www4.wiwiss.fu-berlin.de/bizer/silk/

*$\sim_R$ is known to not hold (i.e. which identify different real world objects).*

Reference links can serve two purposes: Firstly, they can be used to evaluate the quality of a linkage rule. But more importantly, they can also be used to infer a linkage rule which specifies the conditions which must hold true for a pair of entities to be part of $M$:

DEFINITION 3 (LINKAGE RULE). *A linkage rule $l$ assigns a similarity value to each pair of entities:*

$$l: A \times B \to [0,1]$$

*The set of matching entities is given by all pairs for which the similarity according to the linkage rule exceeds a threshold of $0.5$:*

$$M_l = \{(a,b); l(a,b) \geq 0.5, a \in A, b \in B\}$$

In this paper, we consider the problem of learning a linkage rule from a set of reference links:

DEFINITION 4 (LINKAGE RULE LEARNER). *The purpose of a learning algorithm for linkage rules is to learn a linkage rule from a set of reference links:*

$$m: 2^{(A \times B)} \times 2^{(A \times B)} \to (A \times B \to [0,1])$$

The first argument denotes a set of positive reference links, while the second argument denotes a set of negative reference links. The result of the learning algorithm is a linkage rule which should cover as many reference links as possible while generalizing to unknown pairs.

## 3. LINKAGE RULE REPRESENTATION

In this section we introduce an expressive linkage rule representation. We represent a linkage rule as a tree which is built from four basic operators:

**Property Operator:** Retrieves all values of a specific property $p$ of each entity, such as its label property.

**Transformation Operator:** Transforms the values of a set of property or transformation operators $\vec{v}$ according to a specific data transformation function $f^t$. Examples of common transformation functions include case normalization, tokenization and concatenation of values from multiple operators. Multiple transformation operators can be nested in order to apply a sequence of transformations.

**Comparison Operator:** Evaluates the similarity between two entities based on the values of two property or transformation operators $v_a$ and $v_b$ by applying a distance measure $f^d$ and a threshold $\theta$. Examples of common distance measures include Levenshtein, Jaccard, or geographic distance.

**Aggregation Operator:** Due to the fact that, in most cases, the similarity of two entities cannot be determined by evaluating a single comparison, an aggregation operator combines the similarity scores from multiple comparison or aggregation operators $\vec{s}$ into a single score according to a specific aggregation function $f^a$. Examples of common aggregation functions include the weighted average or yielding the minimum score of all operators.

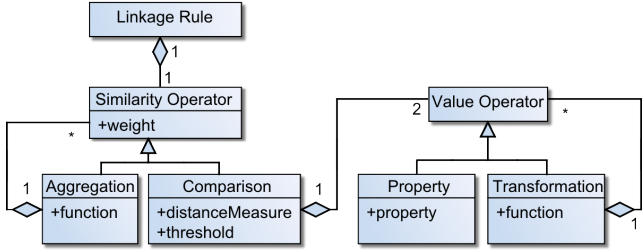

Figure 1: Structure of a linkage rule

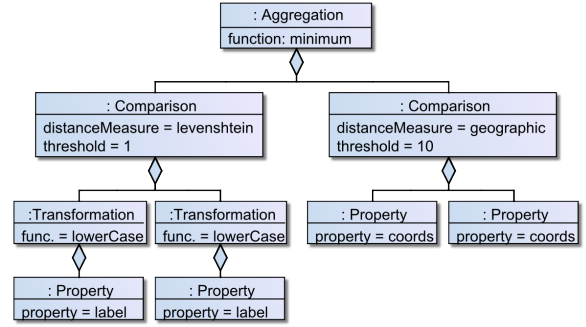

Figure 2: Example linkage rule

The linkage rule tree is strongly typed [26] i.e. only specific combinations of the four basic operators are allowed. Figure 1 specifies the valid structure of a linkage rule.

**Discussion.** Our representation of a linkage rule differs from other representations in record linkage in a number of ways: First of all, each comparison operator accepts two value operators allowing the matching of data sets that are represented using different schemata.

Matching between different schemata is also enabled by the introduction of transformation operators. For example, a data source which uses the FOAF vocabulary [5] may represent person names using the `foaf:firstName` and `foaf:lastName` properties while a data source using the DBpedia ontology may represent the same names using just the `dbpedia:name` property. In order to compare entities expressed in different schemata or data formats, their values have to be normalized prior to comparing them for similarity. In this example we could achieve this in two ways: We could concatenate `foaf:firstName` and `foaf:lastName` into a single name before comparing them to `dbpedia:name` by using a character-based distance measure such as the Levenshtein distance. Alternatively, we could split the values of `dbpedia:name` using a tokenizer and compare them to the values of `foaf:firstName` and `foaf:lastName` by using a token-based distance measure such as the Jaccard coefficient.

Another motivation for transformation operators is the matching of noisy data sets. A common example is data sources which contain values using an inconsistent letter case (e.g. "iPod" vs. "IPOD"). A way to address case inconsistency is to normalize all values to lower case prior to comparing them.

Finally, we allow aggregation operators to be nested which enables us to represent non-linear classifiers beyond pure boolean classifiers. The subsequent related work section will discuss how our representation can be reduced to existing approaches.

It is outside of the scope of this paper to present methods to execute linkage rules which are based on the proposed representation. A method to efficiently execute such linkage rules can be found in [19].

**Example.** Figure 2 shows a simple example of a linkage rule for interlinking cities. In this example, the linkage rule compares the labels as well as the coordinates of the entities. The labels are normalized by converting them to lower case prior to comparing them with the Levenshtein distance while allowing for a maximum distance of 1. The similarity score of the labels is then aggregated with the geographic similarity score into a single score by using the minimum aggregation i.e. both values must exceed the threshold of 0.5 in order to generate a link.

**Semantics.** We now define the semantics of the individual operators: We distinguish between 2 types of operators: *value operators* and *similarity operators*. While value operators provide a function which yields a discriminative value for a single entity, similarity operators provide a function which determines how similar two given entities are.

Given two data sources $A$ and $B$, a *value operator* yields a function which returns a discriminative value for a given entity $e$ by which it can be compared to other entities. Thus, it returns a value from the set:[2]

$$\mathcal{V} := [A \cup B \to \Sigma]$$

Where $\Sigma$ denotes a (possibly empty) set of values.

We now introduce two value operators: *property operators* and *transformation operators*:

DEFINITION 5 (PROPERTY OP.). *A property operator retrieves all values of a specific property of an entity:*

$$v^p : P \to \mathcal{V}$$
$$p \mapsto (e \mapsto e.p)$$

*where $p$ denotes the property to be retrieved by the operator.*

DEFINITION 6 (TRANSFORMATION OP.). *A transformation operator transforms the input values according to a specific data transformation function:*

$$v^t : (\mathcal{V}^* \times \mathcal{F}^t) \to \mathcal{V}$$
$$(\vec{v}, f^t) \mapsto (e \mapsto f^t(v_1(e), v_2(e), \ldots, v_n(e)))$$

*$\vec{v}$ is a vector of operators: $v_1, v_2, \ldots, v_n$. The transformation function $f^t$ may be any function which transforms the value sets provided by the operators into a single value set:*

$$f^t : \Sigma^n \to \Sigma$$

Although in general we do not impose any restriction on the concrete transformation functions which can be used, Table 1 lists the functions which we employed in our experiments. An example of a transformation operator which concatenates the first and the last name of entities about persons is:

$$v^t((v^p(\text{firstName}), v^p(\text{lastName})), concatenate)$$

Note that transformations also may be nested.

---

[2] $[X \to Y]$ denotes $Y^X$ i.e. the space of all functions $X \to Y$

| Transformations | |
|---|---|
| lowerCase | Converts all values to lower case |
| tokenize | Splits all values into tokens |
| stripUriPrefix | Strips the URI prefixes (e.g. `http://dbpedia.org/resource/`) |
| concatenate | Concatenates the values from two value operators |

Table 1: Transformations used in all experiments

| Distance Measures | |
|---|---|
| levenshtein | Levenshtein distance |
| jaccard | Jaccard distance coefficient |
| numeric | The numeric difference |
| geographic | The geographical distance in meters |
| date | Distance between two dates in days |

Table 2: Distance functions used in all experiments

A *similarity operator* returns a function which assigns a value from the interval [0,1] to each pair of entities:

$$\mathcal{S} := [A \times B \to [0,1]]$$

We consider two types of similarity operators: *comparison operators* which compare the result of two value operators and *aggregation operators* which aggregate multiple similarity operators.

DEFINITION 7 (COMPARISON OP.). *Given two value operators $v_a$ and $v_b$, a comparison operator is defined as:*

$$s^c : (\mathcal{V} \times \mathcal{V} \times \mathcal{F}^d \times \mathbb{R}) \to \mathcal{S}$$

$$(v_a, v_b, f^d, \theta) \mapsto \left((e_a, e_b) \mapsto \begin{cases} 1 - \frac{d}{\theta} & \text{if } d \leq \theta \\ 0 & \text{if } d > \theta \end{cases}\right)$$

$$\text{with } d := f^d(v_a(e_a), v_b(e_b))$$

$f^d$ defines the distance measure which is used to compare the values of both value operators:

$$f^d : \Sigma \times \Sigma \to \mathbb{R}$$

Table 2 lists the distance functions which we employed in our experiments. An example of a comparison operator which compares the name of the entities in the first data set with the lower cased labels of the entities in the second data set using the Levenshtein distance is:

$$s^c(v^p(\text{name}), v^t((v^p(\text{label})), lowerCase), levenshtein, 1)$$

DEFINITION 8 (AGGREGATION OP.). *Given a set of similarity operators s an aggregation operator is defined as:*

$$s^a : (\mathcal{S}^* \times \mathbb{N}^* \times \mathcal{F}^a) \to \mathcal{S}$$

$$(\vec{s}, \vec{w}, f^a) \mapsto ((e_a, e_b) \mapsto f^a(s_e, w))$$

$$\text{with } s_e := (s_1(e_a, e_b), s_2(e_a, e_b), \ldots, s_n(e_a, e_b))$$

$\vec{w}$ denotes the weights which are used by the aggregation function $f^a$ to combine the values:

$$f^a : \mathbb{R}^n \times \mathbb{N}^n \to \mathbb{R}$$

The first argument contains the similarity scores returned by the operators of this aggregation while the second argument contains a weight for each of the operators.

| Aggregations | |
|---|---|
| max | $f^t(s,w) := max(s)$ |
| min | $f^t(s,w) := min(s)$ |
| wmean | $f^t(s,w) := \frac{\sum_{i=1}^n w_i s_i}{\sum_{i=1}^n w_i}$ |

Table 3: Aggregation functions used in all experiments

Table 3 lists the aggregation functions which we employed in our experiments. Note that aggregations can be nested i.e. non-linear hierarchies can also be expressed.

## 4. RELATED WORK

Many approaches suitable for learning binary classifiers have been adapted for learning linkage rules [22]. This section introduces the most popular approaches and discusses how they compare to our approach.

**Naive Bayes.** Based on the original Fellegi-Sunter statistical model [15] of record linkage, methods from Bayesian statistics such as *Naive Bayes* classifiers have been used to learn linkage rules [32]. Compared to our approach, Naive Bayes is not capable of expressing data transformations. It has been shown to perform worse than other approaches for entity matching such as support vector machines and decision trees [27].

**Linear Classifiers.** Arasu et. al. [1] categorize widely used approaches for representing linkage rules as *threshold-based boolean classifiers* and *linear classifiers*. Using the introduced representation, we define linear classifiers as:

DEFINITION 9 (LINEAR CLASSIFIER). *Given a vector of comparison operators $\vec{s}$ and a vector of weights $\vec{w}$, a linear classifier is defined as:*

$$s^a(\vec{s}, \vec{w}, wmean) = (e_a, e_b) \mapsto \frac{\sum_{i=1}^n w_i s_i(e_a, e_b)}{\sum_{i=1}^n w_i}$$

Our representation subsumes linear classifiers and extends them in 3 ways:

1. It includes data transformations.
2. It generalizes the aggregation function i.e. allows functions other than *wmean*.
3. It allows aggregations to be nested.

The most popular method to model linear classifiers are *support vector machines* (SVM) [7]. A SVM is a binary linear classifier which maps the input variables into a high-dimensional space where the two classes are separated by a hyperplane via a kernel function [3]. One popular application of SVMs to entity matching is MARLIN (Multiply Adaptive Record Linkage with INduction) [2], which uses SVMs to learn linear classifiers. While SVMs can be extended to model non-linear classifiers, they are not suitable to learn data transformations.

**Threshold-based Boolean Classifiers.** We first define threshold-based boolean classifiers:

DEFINITION 10 (THRESHOLD-BASED BOOLEAN CL.). *Given a vector of comparison operators $\vec{s}$, where each comparison operator compares a pair of properties using a specific distance measure $f^d$ and a threshold $\theta$, a threshold-based boolean classifiers is defined as:*

$$s^a(\vec{s}, min) = (e_a, e_b) \mapsto min\{s_i(e_a, e_b) : 1 < i \leq n\}$$

Note that using the minimum function is equivalent to the conjunction of all comparisons.

Our representation subsumes threshold-based boolean classifiers and extends them in 2 ways:

1. It includes data transformations.
2. It generalizes the aggregation function i.e. allows functions other than $min$ and $max$.

In literature, threshold-based boolean classifiers are usually represented with decision trees. A major advantage of decision trees is that they provide explanations for each classification and thus can be understood and improved manually. Active Atlas [28, 29] learns mapping rules consisting of a combination of predefined transformations and similarity measures. TAILOR [11] is another tool which employs decision trees to learn linkage rules.

**Genetic Programming.** Genetic programming (GP) is an extension of the genetic algorithm [17] which has been first proposed, in tree-based form, by Cramer [8]. As genetic programming represents candidate solutions as trees, linkage rules can be directly represented.

Genetic programming algorithms usually start with a random population and evolve the population using three common genetic operations [24]:

**Reproduction** copies an individual without modification.
**Crossover** recombines two individuals.
**Mutation** applies a random modification to an individual.

These operations are applied to individuals in the population which have been selected based on a fitness measure which determines how close a specific individual is to the desired solution. The evolution of the population stops as soon as either the configured maximum number of iterations or the maximum F-measure is reached.

To the best of our knowledge, genetic programming for learning linkage rules has only been applied by Carvalho et. al. so far [9, 6, 10]. Their approach uses genetic programming to learn how to combine a set of presupplied pairs of the form `<attribute, similarity function>` (e.g. `<name, Jaro>`) into a linkage rule. These pairs can be combined by the genetic programming method to a linkage rule tree by using mathematical functions (e.g. +, -, *, /, exp) and constants. Carvalho et. al. show that their method produces better results as the state-of-the-art SVM based approach by MARLIN [10]. Their approach is very expressive although it cannot express data transformations. On the downside, using mathematical functions to combine the similarity measures does not fit any commonly used linkage rule model [14] and leads to complex and difficult to understand linkage rules.

## 5. APPROACH

This section describes the GenLink approach in detail. The pseudocode of GenLink is given in Algorithm 1.

The algorithm starts by generating an initial population of candidate linkage rules according to the method described in Section 5.1.

After the initial population has been generated it is iteratively evolved. In each iteration, a new population is generated by creating new linkage rules from the existing population until the population size is reached.

**Algorithm 1** Pseudocode of the GenLink algorithm. The specific parameter values used in our experiments are listed in Section 6.1.

```
P ← generate initial population
while(maximum iterations nor full F−measure reached) {
  P' ← ∅
  while(|P'| < populationsize) {
    r_1, r_2 ← select two linkage rules from P
    op ← select random crossover operator
    p ← random number from interval [0,1]
    if(p < mutationprobability) {
      r_r ← generate random linkage rule
      P' ← P' ∪ op(r_1, r_r)
    } else {
      P' ← P' ∪ op(r_1, r_2)
    }
  }
  P ← P'
}
return best linkage rule from P
```

A new linkage rule is generated according to the following steps: First, two linkage rules are selected from the population according to the selection method described in Section 5.2. In addition, a random crossover operator is selected from the set of available operators. The basic idea of our approach is to provide a specific crossover operator for each aspect of the linkage rule. For instance, the threshold crossover operator only modifies the threshold of the specific comparison while the transformation crossover operator combines the transformations of both linkage rules. The set of provided crossover operators is described in Section 5.3. The selected operator is used to either mutate one of the selected linkage rules or to combine both linkage rules into a new linkage rule.

The algorithm stops when either a predefined number of iterations is reached or when one linkage rule in the population reaches the full F-measure. The best linkage rule in the final population is returned by the algorithm.

### 5.1 Generating the Initial Population

In genetic programming the initial population is usually generated randomly. Previous work has shown that starting with a fully random population works well on some record linkage data sets [6]. Two circumstances increase the search space (i.e. the set of all possible linkage rules) considerably: Firstly, data sets with a high number of properties. Secondly, if data sets which are represented using different schemata are to be matched the search space includes all possible property pairs from the source and target data set. In order to reduce the size of the search space, we employ a simple algorithm which preselects property pairs which hold similar values: Before the population is generated, we build a list of property pairs which hold similar values as described below. Based on that, random linkage rules are built by selecting property pairs from the list and building a tree by combining random data transformations, comparisons and aggregations.

**Finding Compatible Properties.** The purpose of this step is to generate a list of pairs of properties which hold similar values. For each possible property pair, the values of the entities referenced by a positive reference link are analyzed. This is done by tokenizing and lowercasing the values and generating a new property pair of the form $(p1, p2, measure)$ if there is a distance measure in a provided list of functions according to which 2 tokens are similar given a certain threshold $\theta_d$. In our experiments, we only used the levensthein distance with a threshold of 1. The pseudocode is given in Algorithm 2.

**Algorithm 2** Find compatible properties given a set of reference links $R^+$ and a distance threshold $\theta$

pairs $\leftarrow \emptyset$
**for** all $(e_a, e_b) \in R^+$ {
　**for** all properties $e_a.pi$ and $e_b.pj$ {
　　**for** all distance functions $f^d$ {
　　　$v_a \leftarrow tokenize(lowerCase(e_a.pi))$
　　　$v_b \leftarrow tokenize(lowerCase(e_b.pj))$
　　　**if** $(f^d(v_a, v_b) < \theta_d)$ add $(pi, pj, f^d)$ to pairs
}}}
**return** pairs

Figure 3 illustrates a simple example with two entities. In this example, the following two property pairs are generated: $(label, label, levensthein)$ and $(point, coord, geographic)$.

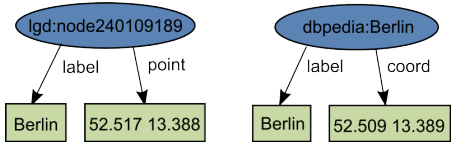

**Figure 3: Finding compatible properties**

**Generating a Random Linkage Rule.** A random linkage rule is generated according to the following rules: First of all, a linkage rule is built consisting of a random aggregation and up to two comparisons. For each comparison a random pair from the pre-generated list of compatible properties is selected. In addition, with a possibility of 50%, a random transformation is appended to each property.

Note that although the initial linkage rule trees are very small, this does not limit the algorithm from growing bigger trees by using the genetic operators.

### 5.2 Selection

Starting with the initial population, the genetic algorithm breeds a new population by evolving selected linkage rules using the genetic operations. The linkage rules are selected from the population based on two functions: The *fitness function* and the *selection method*.

The purpose of the **fitness function** is to assign a value to each linkage rule which indicates how close the given linkage rule is to the desired solution. A disadvantage of using the F-measure as fitness function is that it may yield skewed results if the number of positive and negative reference links is unbalanced as it only takes the true negative rate into account. We use *Matthews correlation coefficient* (MCC) as fitness measure. Matthews correlation coefficient [25] is defined as the degree of the correlation between the actual and predicted classes:

$$\text{MCC} = \frac{n_{tp} \times n_{tn} - n_{fp} \times n_{fn}}{\sqrt{(n_{tp} + n_{fp})(n_{tp} + n_{fn})(n_{tn} + n_{fp})(n_{tn} + n_{fn})}}$$

$n_{tp}$, $n_{tn}$, $n_{fp}$ and $n_{fn}$ denote the number of true positives, true negatives, false positives and false negatives which are computed based on the provided reference links (ignoring the remaining part of the data set). In order to prevent linkage rules from growing indefinitely, we penalize linkage rules based on their number of operators: $fitness = mcc - 0.05 * operatorcount$.

Based on the fitness of each linkage rule, the **selection method** selects the linkage rules to be evolved. As selection method we chose tournament selection as it has been shown to produce strong results in a variety of GP systems [23] and is easy to parallelize.

### 5.3 Crossover Operators

Instead of using subtree crossover, which is commonly used in genetic programming, we use a set of specific crossover operators which are tailored to the structure of a linkage rule. For each crossover operation, an operator from this set is selected randomly and applied to two selected linkage rules. We reduce mutation to a crossover operation with a randomly generated new linkage rule. This is known as *headless chicken crossover*. If the root node is selected as the crossover point in the first linkage rule this results in the complete replacement of the given linkage rule with a randomly generated linkage rule.

Each operator learns one aspect of the linkage rule. The contribution of the proposed operators to the learning performance over subtree crossover is evaluated experimentally in Section 6.3.

For evolving linkage rules, we propose the following operators:

**Function Crossover.** This crossover operator is used to find the best distance, transformation or aggregation function. Function crossover selects one operator at random in each linkage rule and interchanges the functions. For example, it may select a comparison with the Levenshtein distance function in the first linkage rule and a comparison with the Jaccard distance function in the second linkage rule and then interchanges these two functions. The pseudocode for function crossover is given in Algorithm 3.

**Algorithm 3** Function Crossover

**def** cross $(r1: \text{Rule}, r2: \text{Rule}) = \{$
　$nodeType \leftarrow$ select random type from {Transformation,
　　↪ Comparison, Aggregation}
　$cmp1 \leftarrow$ random node of $nodeType$ from $r1$
　$cmp2 \leftarrow$ random node of $nodeType$ from $r2$

　**return** $r1$ with $cmp1.f^d \leftarrow cmp2.f^d$
$\}$

**Operators Crossover.** As a linkage rule usually needs to combine multiple comparisons, this operator combines aggregations from both linkage rules. For this, it selects two aggregations, one from each linkage rule and combines theirs comparisons. The comparisons are combined by selecting all comparisons from both aggregations and removing each comparison with a probability of 50%. For example, it may select an aggregation of a label comparison and a date comparison in the first linkage rule and an aggregation of a label comparison and a comparison of the geographic coordinates in the second linkage rule. In this case the operator replaces the selected aggregations with a new aggregation which contains all 4 comparisons and then removes each comparison with a probability of 50%. Note that the comparisons are exchanged including the complete subtree i.e. the distance functions as well as existing transformations are retained. The pseudocode is given in Algorithm 4. Figure 4 illustrates

**Algorithm 4** Operators Crossover

**def** cross ($r1$: Rule, $r2$: Rule) = {
  $agg1 \leftarrow$ random aggregation from $r1$
  $agg2 \leftarrow$ random aggregation from $r2$

  $ops \leftarrow \emptyset$
  **for** all operators $o$ in $agg1$ and $agg2$ {
    $p \leftarrow$ random number from interval [0,1]
    **if** ($p > 0.5$)
      add $o$ to $ops$
  }

  **return** $r1$ with $agg1.operators \leftarrow ops$
}

a simple operators crossover on two linkage rules.

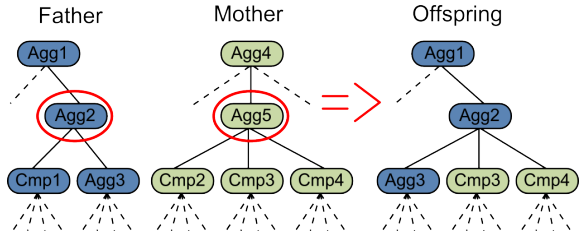

Figure 4: Operators Crossover

**Aggregation Crossover.** While for some data sets it is sufficient to use pure linear or boolean classifiers, for other data sets the accuracy can be improved by allowing non-linear aggregations (see Section 6.3). In order to learn aggregation hierarchies, the aggregation crossover operator selects a random aggregation or comparison operator in the first linkage rule and replaces it with a random aggregation or comparison operator from the second linkage rule. This way, the operator builds a hierarchy as it may select operators from different levels in the tree. For example, it may select a comparison in the first linkage rule and replace it with a aggregation of multiple comparisons from the second linkage rule. Note that Aggregation Crossover is similar to subtree crossover but only operates on aggregation and comparison nodes. The pseudocode is given in Algorithm 5. Figure 5 illustrates a simple aggregation crossover on two

**Algorithm 5** Aggregation Crossover

**def** cross ($r1$: Rule, $r2$: Rule) = {
  $o1 \leftarrow$ random aggregation or comparison from $r1$
  $o2 \leftarrow$ random aggregation or comparison from $r2$

  **return** $r1$ with $o1$ replaced by $o2$
}

linkage rules.

**Transformation Crossover.** This operator is used to recombine the transformations of two linkage rules. By combining the transformation operators of both linkage rules,

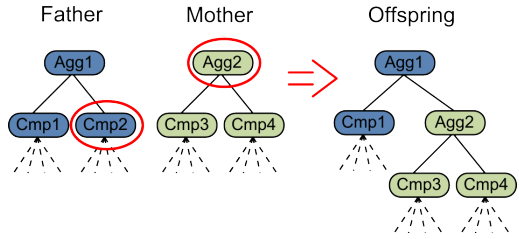

Figure 5: Aggregation Crossover

it can build up chains of transformations. In both linkage rules, transformation crossover randomly selects an upper and a lower transformation operator. The next step is to recombine the paths between the upper and the lower transformation by executing a two point crossover. Finally, duplicated transformations are removed. The pseudocode is given in Algorithm 6. Figure 6 illustrates a transformation

**Algorithm 6** Transformation Crossover

**def** cross ($r1$: Rule, $r2$: Rule) = {
  $t1_{upper}, t1_{lower} \leftarrow$ random transformations from $r1$
  $t2_{upper}, t2_{lower} \leftarrow$ random transformations from $r2$

  **return** $r1$ with:
    $t1_{upper}$ replaced by $t2_{upper}$
    $t2_{lower}.\vec{v}$ replaced by $t1_{lower}.\vec{v}$
}

crossover. In this example, the `tokenize` operator was se-lected as both upper and lower transformation in the first linkage rule. In the second linkage rule, the `tokenize` operator was selected as upper transformation while the `stem` operator was selected as lower transformation. The `tokenize` operator in the first linkage rule is then replaced by the path between the upper and the lower transformation in the second linkage rule.

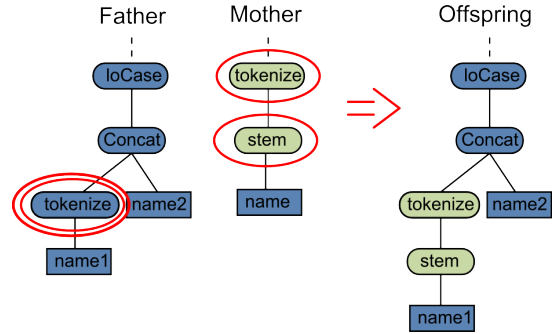

Figure 6: Transformation Crossover

**Threshold Crossover.** This crossover operator combines the distance thresholds of both linkage rules. For this, one comparison operator is selected at random in each linkage rule. The new threshold is then set to the average of both comparisons. The pseudocode is given in Algorithm 7.

**Weight Crossover.** This crossover operator combines the weights of both linkage rules analogous to the threshold

**Algorithm 7** Threshold Crossover

```
def cross (r1: Rule, r2: Rule) = {
  agg1 ← random comparison from r1
  agg2 ← random comparison from r2

  return r1 with (agg1.θ ← 0.5 · (agg1.θ + agg2.θ))
}
```

| Parameter | Value |
|---|---|
| Population size | 500 |
| Maximum iterations | 50 |
| Selection method | Tournament selection |
| Tournament size | 5 |
| Probability of Crossover | 75% |
| Probability of Mutation | 25% |
| Stop Condition | F-measure = 1.0 |

Table 4: Parameters

crossover. For this, it selects a comparison or aggregation operator in each linkage rule and updates the weight in the first operator to the average of both.

## 6. EVALUATION

In this section we evaluate our approach experimentally: Section 6.1 describes the experimental setup. The overall learning results for several real world data sets are presented in Section 6.2. Finally, Section 6.3 evaluates the contribution of specific parts of our algorithm to the accuracy of the learned linkage rules.

### 6.1 Experimental Setup

The GenLink algorithm has been implemented in the Silk Link Discovery Framework which can be downloaded from the project homepage[3]. The Silk Link Discovery Framework supports users in discovering relationships between data items within different Linked Data sources. All experiments have been executed using Version 2.5.3 of the Silk Link Discovery Framework.

Because genetic algorithms are non-deterministic and may yield different results in each run, all experiments have been run 10 times. For each run the reference links have been randomly split into 2 folds for cross-validation. The results of all runs have been averaged and the standard deviation has been computed. For each experiment, we provide the evaluation results with respect to the training set as well as the validation set. All experiments have been run on a 3GHz Intel(R) Core i7 CPU with 4 cores while the Java heap space has been restricted to 1GB.

**Parameters.** Table 4 lists the parameters which have been used in all experiments. As it is the purpose of the developed method to work on arbitrary data sets without the need to tailor its parameters to the specific data sets that should be matched, the same parameters have been used for all experiments.

**Data sets.** For evaluation, we used six data sets from three areas: (1) We evaluate the learning performance on two well-known record linkage data sets and compare the performance with an existing state-of-the-art genetic programming approach. (2) We evaluate our approach with two data sets from the Ontology Alignment Evaluation Initiative[4] and compare our results to the participating systems. (3) We compare the learned linkage rules with linkage rules created by a human expert for two data sets.

[3] http://www4.wiwiss.fu-berlin.de/bizer/silk/

|  | Entities | | Reference Links | |
|---|---|---|---|---|
|  | $\|A\|$ | $\|B\|$ | $\|R_+\|$ | $\|R_-\|$ |
| Cora | 1879 | | 1617 | 1617 |
| Restaurant | 864 | | 112 | 112 |
| SiderDrugbank | 924 | 4772 | 859 | 859 |
| NYT | 5620 | 1819 | 1920 | 1920 |
| LinkedMDB | 199 | 174 | 100 | 100 |
| DBpediaDrugbank | 4854 | 4772 | 1403 | 1403 |

Table 5: The number of entities in each data set as well as the number of reference links.

|  | Properties | | Coverage | |
|---|---|---|---|---|
|  | $\|A.P\|$ | $\|B.P\|$ | $C_A$ | $C_B$ |
| Cora | 4 | | 0.8 | |
| Restaurant | 5 | | 1.0 | |
| SiderDrugbank | 8 | 79 | 1.0 | 0.5 |
| NYT | 38 | 110 | 0.3 | 0.2 |
| LinkedMDB | 100 | 46 | 0.4 | 0.4 |
| DBpediaDrugbank | 110 | 79 | 0.3 | 0.5 |

Table 6: The total number of properties in each data set as well as the percentage of properties which are actually set on an entity.

The first two data sets are frequently-used record linkage data sets while the following 4 sets are RDF data sets. While the record linkage data sets are already adhering to a consistent schema, the RDF data sets are split into a source and a target data set which adhere to different schemata.

Table 5 lists the used data sets together with the number of entities as well as the number of reference links in each data set. As only positive reference links have been provided by the data set providers, we generated the negative reference links. For two positive links $(a,b) \in R^+$ and $(c,d) \in R^+$ we generated two negative links $(a,d) \in R^-$ and $(c,b) \in R^-$. For the Cora and Restaurant data set this is sound as the provided positive links are complete. Since the remaining data sources are split into source and target data sets, generating negative reference links is possible as entities in the source and target data sets are internally unique.

Table 6 shows the number of properties in the source and target data sets and their coverage i.e. the percentage of properties which are actually set on an entity on average. The following section will provide more detailed information for each data set.

### 6.2 Overall Results

In this section we compare the overall performance of GenLink on all 6 data sets.

*Frequently Used Record Linkage Datasets*

A number of data sets have been used frequently to evaluate the performance of different record linkage approaches. Following this practice, we compared the overall learning performance of our approach with the approach proposed by Carvalho et. al. [10]. In their experiments, Carvalho et. al. report to produce better results than the state-of-the-art SVM based approach by MARLIN. The related work section

[4] http://oaei.ontologymatching.org

provides more details about how their approach compares to ours technically.

We chose 2 data sets which are also used by Carvalho et. al.: the *Cora* data set and *Restaurant* data set. The **Cora** data set contains citations to research papers from the Cora Computer Science research paper search engine. For each citations it contains the title, the author, the venue as well as the date of publication. The **Restaurant** data set contains a set of records from the Fodor's and Zagat's restaurant guides. For each restaurant it contains the name, address, phone number as well as the type of restaurant. For both data sets, we used the XML version[5] which is provided by Draisbach et. al.

Table 7 summarizes the cross validation results for the Cora data set. On average, our approach achieved an F-

| Iter. | Time in s ($\sigma$) | Train. F1 ($\sigma$) | Val. F1 ($\sigma$) |
|---|---|---|---|
| 0 | 5.5 (0.7) | 0.880 (0.030) | 0.877 (0.031) |
| 10 | 28.6 (2.7) | 0.949 (0.018) | 0.945 (0.021) |
| 20 | 60.1 (4.1) | 0.965 (0.005) | 0.962 (0.005) |
| 30 | 93.6 (6.1) | 0.968 (0.003) | 0.965 (0.004) |
| 40 | 129.4 (9.7) | 0.968 (0.002) | 0.965 (0.004) |
| 50 | 185.8 (26.7) | 0.969 (0.003) | 0.966 (0.004) |
| Ref. | - | 0.900 (0.010) | 0.910 (0.010) |

**Table 7: Results for the Cora data set. The last row contains the best results of Carvalho et. al. for comparison.**

measure of 96.9% against the training set and 96.6% against the validation set and needed about 3 minutes to perform all 50 iterations on the test machine. The learned linkage rules compared by title, author and venue. Figure 7 shows an example of a learned linkage rule which reached the top F-measure. For the same data set, Carvalho et. al. report an F-measure of 90.0% against the training set and 91.0% against the validation set [10]. We suspected the main reason for the better performance of our method on this data set to be found in the inclusion of data transformations in our learning approach. To confirm this claim we re-executed our method with one limitation: No data transformations were allowed to be used in a linkage rule. With this limitation, the performance of our methods declined to an F-measure of 91.2% against the training set and 90.5% against the validation set approximately matching the numbers of Carvalho et. al.. Figure 8 shows a learned linkage rule without transformations.

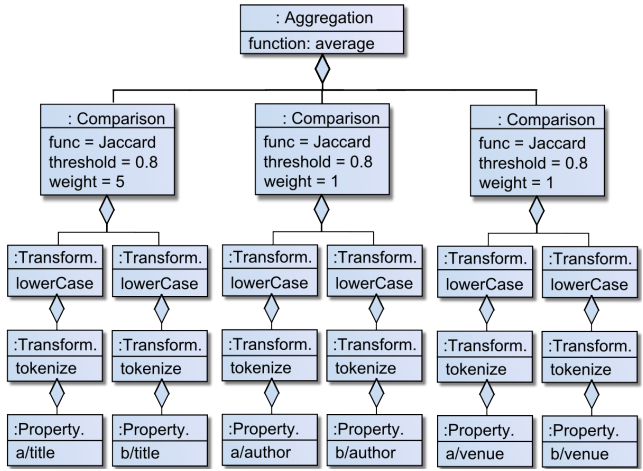

**Figure 7: Cora: Learned linkage rule**

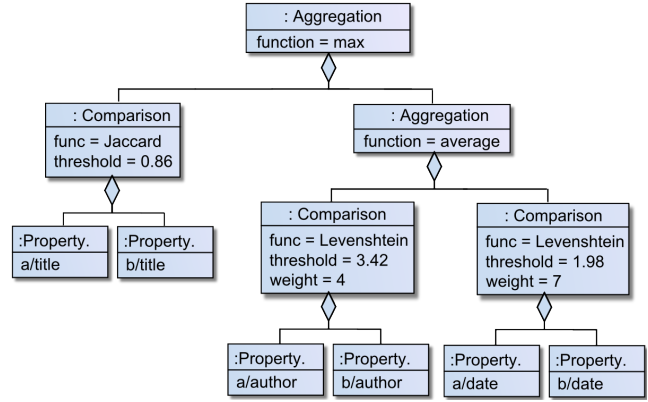

**Figure 8: Cora: Learned linkage rule without transformations**

Table 8 summarizes the cross validation results for the Restaurant data set. On average, our approach achieved an F-measure of 99.6% against the training set and 99.3% against the validation set. For the same data set, Carvalho et. al. report an F-measure of 100.0% against the training set, but only 98.0% against the validation set [10].

| Iter. | Time in s ($\sigma$) | Train. F1 ($\sigma$) | Val. F1 ($\sigma$) |
|---|---|---|---|
| 0 | 0.4 (0.1) | 0.953 (0.038) | 0.951 (0.039) |
| 10 | 2.0 (0.9) | 0.996 (0.004) | 0.992 (0.006) |
| 20 | 3.1 (1.9) | 0.996 (0.004) | 0.993 (0.006) |
| 30 | 4.1 (3.0) | 0.996 (0.004) | 0.993 (0.006) |
| 40 | 5.2 (4.0) | 0.996 (0.004) | 0.993 (0.006) |
| 50 | 6.3 (5.3) | 0.996 (0.004) | 0.993 (0.006) |
| Ref. | - | 1.000 (0.000) | 0.980 (0.010) |

**Table 8: Results for the Restaurant data set. The last row contains the best results of Carvalho et. al. for comparison.**

*Ontology Alignment Evaluation Initiative*

The *Ontology Alignment Evaluation Initiative* (OAEI) is an international initiative aimed at organizing the evaluation of different ontology matching systems. In addition to schema matching, OAEI also includes an instance matching track since 2009 which regularly evaluates the ability to identify similar entities among different Linked Data sources.

The **SiderDrugBank** data set was selected from the OAEI 2010 data interlinking track[6]. We chose this data set amongst the other drug related data sets because it was the one for which the participating systems ObjectCoref [18] and RiMOM [30] performed the worst. This data set contains drugs from Sider, a data set of marketed drugs and their side effects, and DrugBank, containing drugs approved by the US Federal Drugs Agency. Positive reference links are provided by the OAEI.

---

[5]http://www.hpi.uni-potsdam.de/naumann/projekte/dude_duplicate_detection.html

[6]http://oaei.ontologymatching.org/2010/im/index.html

1646

The **NYT** data set was selected from the OAEI 2011 data interlinking track[7]. Amongst the 7 data sets from this track, we chose the data set for which the participating systems performed the worst on average: Interlinking locations in the New York Times data set with their equivalent in DBpedia. Besides other types of entities, the New York Times data set contains 1920 manually curated locations where for each location a link to the same location in DBpedia is set.

For both data sets, we compared our results with the results of the participating systems in the instance matching track of the OAEI. In the OAEI, the systems where asked to identify similar entities in a data set without being allowed to employ existing reference links for matching. Note that as the OAEI only compares unsupervised systems and does not consider supervised systems (i.e. systems which are supplied with existing reference links), our approach has an advantage over the participating systems. For that reason, we used the official OAEI results merely as a baseline for our approach.

Table 9 summarizes the cross validation results for the SiderDrugBank data set. After 30 iterations, our approach reached an F-measure of 97.2% for the training set and 97.0% for the validation set. ObjectCoref and RiMOM achieved only around 50% percent which indicates the difficulty in matching this data set.

| Iter. | Time in s ($\sigma$) | Train. F1 ($\sigma$) | Val. F1 ($\sigma$) |
|---|---|---|---|
| 0  | 21.7 (0.3)    | 0.840 (0.018) | 0.837 (0.018) |
| 10 | 38.8 (2.4)    | 0.943 (0.025) | 0.939 (0.030) |
| 20 | 83.1 (11.1)   | 0.970 (0.007) | 0.969 (0.008) |
| 30 | 147.2 (20.9)  | 0.972 (0.006) | 0.970 (0.007) |
| 40 | 215.6 (28.0)  | 0.972 (0.006) | 0.970 (0.007) |
| 50 | 301.5 (39.0)  | 0.972 (0.006) | 0.970 (0.007) |
| Reference System | | | F1 |
| ObjectCoref | | | 0.464 [18] |
| RiMOM | | | 0.504 [30] |

Table 9: Results for the SiderDrugBank data set.

Table 10 summarizes the cross validation results for the NYT data set. After 50 iterations, our approach reached an F-measure of 97.7% for the training set and 97.4% for the validation set. With an F-measure of 92%, Zhishi.links achieved the best result of the participating systems.

| Iter. | Time in s ($\sigma$) | Train. F1 ($\sigma$) | Val. F1 ($\sigma$) |
|---|---|---|---|
| 0  | 85.2 (1.7)    | 0.703 (0.064) | 0.709 (0.048) |
| 1  | 107.7 (11.0)  | 0.803 (0.037) | 0.803 (0.036) |
| 5  | 260.7 (76.8)  | 0.844 (0.048) | 0.846 (0.048) |
| 10 | 344.5 (86.2)  | 0.854 (0.052) | 0.854 (0.053) |
| 20 | 496.7 (95.0)  | 0.907 (0.074) | 0.906 (0.074) |
| 30 | 652.8 (108.1) | 0.927 (0.069) | 0.928 (0.067) |
| 40 | 804.8 (132.4) | 0.965 (0.039) | 0.963 (0.041) |
| 50 | 975.4 (141.1) | 0.977 (0.024) | 0.974 (0.026) |
| Reference System | | | F1 [13] |
| AgreementMaker | | | 0.69 |
| SEREMI | | | 0.68 |
| Zhishi.links | | | 0.92 |

Table 10: Results for the NYT data set.

---

[7] http://oaei.ontologymatching.org/2011/instance/index.html

*Comparison With Manually Created Linkage Rules*

In addition, we evaluated how the learned linkage rules compare to linkage rules which have been manually created by a human expert for the same data set. For this we selected 2 data sets: A data set about movies and a complex life science data set.

**LinkedMDB:** An easy to understand data set about movies which is non-trivial as the linkage rule cannot just compare by label (different movies may have the same name), but also needs to include other properties such as the date or the director. For evaluation we used a manually created set of 100 positive and 100 negative reference links. Special care was taken to include relevant corner cases such as movies which share the same title but have been produced in different years.

Table 11 summarizes the cross validation results for the LinkedMDB data set.

| Iter. | Time in s ($\sigma$) | Train. F1 ($\sigma$) | Val. F1 ($\sigma$) |
|---|---|---|---|
| 1  | 4.2 (1.2)    | 0.981 (0.011) | 0.959 (0.023) |
| 10 | 19.6 (2.1)   | 0.998 (0.001) | 0.921 (0.007) |
| 20 | 40.3 (3.3)   | 1.000 (0.000) | 0.974 (0.004) |
| 30 | 59.2 (4.2)   | 1.000 (0.000) | 0.999 (0.002) |
| 40 | 74.2 (7.2)   | 1.000 (0.000) | 0.999 (0.002) |
| 50 | 99.7 (12.8)  | 1.000 (0.000) | 0.999 (0.002) |

Table 11: Results for the LinkedMDB data set.

In all runs, the learning algorithm needed no more than 12 iterations in order to achieve the full training F-measure. The learning algorithm learned linkage rules which compare movies by their title and their release date just as the original human-created linkage rule did.

**DBpediaDrugBank:** While the vast majority of linkage rules commonly used in the Linked Data context are very simple, a few of them employ more complex structures. Interlinking drugs in DBpedia and DrugBank is an example where the original linkage rule which has been produced by humans is very complex. In order to match two drugs, it compares the drug names and their synonyms as well as a list of well-known and used identifiers (e.g. the CAS number[8]) which are provided by both data sets but are missing for many entities. In total, the manually written linkage rule uses 13 comparisons and 33 transformations. This includes complex transformations such as replacing specific parts of the strings. All 1,403 links which have been generated by executing the original linkage rule have been used as positive reference links.

Table 12 summarizes the cross validation results for the DBpediaDrugBank data set. The learned linkage rules yield an F-Measure of 99.8% for the training data and 99.4% for the validation data. From the 30th iteration the generated linkage rules on average only use 5.6 comparisons and 3.2 transformations and the parsimony pressure successfully avoids bloating in the subsequent iterations. Thus, the learned linkage rules use less than half of the comparisons and only one-tenth of the transformations of the human written linkage rules.

## 6.3 Detailed Evaluation

While the previous section focused on the evaluation of the performance of the overall algorithm this section focuses on evaluating specific parts of our approach. One of the main claims of this paper is that using the expressive linkage rule representation presented in Section 3 allows the algorithm to learn rules with higher accuracy. We evaluate this claim by comparing the performance of learned linkage rules using the proposed representation versus common representations in record linkage. After that, we show how our approach to generate the initial population improves the average accuracy of the initial linkage rules. Finally, we evaluate how the proposed specialized crossover operators improves the learning performance over subtree crossover.

---

[8] A unique numerical identifier assigned by the "Chemical Abstracts Service"

| Iter. | Time in s ($\sigma$) | Train. F1 ($\sigma$) | Val. F1 ($\sigma$) |
|---|---|---|---|
| 1 | 67.5 (2.2) | 0.929 (0.026) | 0.928 (0.029) |
| 10 | 334.1 (157.4) | 0.994 (0.002) | 0.991 (0.003) |
| 20 | 1014.1 (496.8) | 0.996 (0.001) | 0.988 (0.010) |
| 30 | 1829.7 (919.3) | 0.997 (0.001) | 0.985 (0.016) |
| 40 | 2685.4 (1318.9) | 0.998 (0.001) | 0.994 (0.002) |
| 50 | 3222.2 (1577.7) | 0.998 (0.001) | 0.994 (0.002) |

Table 12: Results for the DBpediaDrugBank data set.

|  | Boolean | Linear | Nonlin. | Full |
|---|---|---|---|---|
| Cora | 0.900 | 0.896 | 0.898 | 0.965 |
| Restaurant | 0.954 | 0.959 | 0.951 | 0.992 |
| SiderDrugBank | 0.931 | 0.956 | 0.966 | 0.970 |
| NYT | 0.714 | 0.716 | 0.724 | 0.916 |
| LinkedMDB | 0.973 | 0.986 | 0.987 | 0.997 |
| DBpediaDrugBank | 0.990 | 0.981 | 0.991 | 0.993 |

Table 13: Representations: F-measure in round 25

*Comparison With Other Linkage Rule Representations*

We presented an approach which uses a linkage rule representation which is more expressive than other representations used in record linkage. Our model includes chains of transformations and is also able to represent non-linear classifiers.

In order to measure the effect of this extended representation, we evaluated the learning performance of 4 representations:

- **Boolean:** Boolean classifiers without transformations
- **Linear:** Linear classifiers without transformations
- **Non-linear:** Non-linear classifiers without transformations
- **Full:** Our Approach with full expressivity

Table 13 shows the F-measure on the validation set after 25 iterations. We now review how the introduction of non-linearity and transformations in our approach improves the learning performance: (1) For the record linkage data sets Cora and Restaurant, the introduction of non-linearity did not improve the learning performance. But, as these data sets are generally rather noisy, the introduction of transformations improved the performance considerably. (2) For the Linked Data sources the non-linear classifiers yield better results than either boolean or linear classifiers. The use of transformations further improves the performance.

*Seeding*

In this experiment we evaluated if our approach of generating the initial population improves over the completely random generation of linkage rules which is usually used in genetic programming. Table 14 compares the average F-measure of the linkage rules in the initial population for each data set. The table shows that for data sets with only a few properties, such as the Cora and the Restaurant data set, the seeding does not yield a significant improvement. However, for data sets with many properties, the improvement over the complete random generation is significant and increases the average F-measure of the linkage rules in the population considerably.

|  | Random | Seeded |
|---|---|---|
| Cora | 0.849 (0.045) | 0.865 (0.018) |
| Restaurant | 0.963 (0.010) | 0.985 (0.012) |
| SiderDrugBank | 0.624 (0.181) | 0.848 (0.013) |
| NYT | 0.178 (0.164) | 0.701 (0.072) |
| LinkedMDB | 0.719 (0.175) | 0.975 (0.008) |
| DBpediaDrugBank | 0.702 (0.217) | 0.957 (0.013) |

Table 14: Seeding: Initial F-measure

| 10 Iterations | Subtree C. | Our Approach |
|---|---|---|
| Cora | 0.943 (0.015) | 0.951 (0.013) |
| Restaurant | 0.997 (0.004) | 0.997 (0.004) |
| SiderDrugBank | 0.919 (0.013) | 0.963 (0.013) |
| NYT | 0.814 (0.015) | 0.834 (0.016) |
| LinkedMDB | 0.985 (0.012) | 0.991 (0.009) |
| DBpediaDrugBank | 0.992 (0.002) | 0.994 (0.002) |
| 25 Iterations | Subtree C. | Our Approach |
| Cora | 0.959 (0.007) | 0.967 (0.003) |
| Restaurant | 0.997 (0.004) | 0.997 (0.004) |
| SiderDrugBank | 0.974 (0.004) | 0.987 (0.003) |
| NYT | 0.814 (0.005) | 0.916 (0.006) |
| LinkedMDB | 0.996 (0.007) | 0.998 (0.003) |
| DBpediaDrugBank | 0.994 (0.001) | 0.997 (0.002) |

Table 15: Crossover experiment

*Crossover Operators*

Subtree crossover is the de-facto standard in genetic programming. We evaluated the actual contribution to the learning performance of using specialized crossover operators instead as described in Section 5.3. Table 15 compares the performance of both configurations after executing 10 iterations and again after 25 iterations.

In all data sets our approach either matches the subtree crossover results or outperforms them. In addition, the specialized crossover operators in our approach have the advantage that each operator only covers a specific aspect of a linkage rule. Thus, the operators can be selectively enabled to control which aspects of a linkage rule are learned.

## 7. CONCLUSION

We presented the GenLink algorithm for learning expressive linkage rules using genetic programming. GenLink employs a linkage rule representation which is more expressive than previous work and is able to represent non-linear rules and may include data transformations which normalize the values prior to comparison. The experimental evaluation shows that this extended representation allows GenLink to generate linkage rules with a higher accuracy as could be achieved with boolean and linear linkage rules. It further shows that the proposed algorithm outperforms the state-of-the-art genetic programming approach for record linkage presented by Carvalho et. al. [10].

We have implemented the GenLink algorithm as part of the Silk Link Discovery Framework. The algorithm can thus directly be used by Linked Data publishers and consumers to set RDF links pointing into other data sources.